\documentclass[12pt,a4paper]{article}
\pdfoutput=1
\usepackage[euler]{textgreek}
\usepackage[format=hang]{caption}
\usepackage{epsfig,amssymb,amsmath,graphicx,subcaption,verbatim,hyperref,xcolor,ulem,epstopdf,psfrag,pstool,braket,array,enumerate,multirow,wrapfig}
\usepackage{graphics,color}
\newdimen\savedimen

\numberwithin{equation}{section}

\begin{document}

\thispagestyle{empty}
\begin{center}
	{{\bf \Large Electromagnetic transition rates of Carbon-12 and Oxygen-16 in rotational-vibrational models }} 
	\\[8mm]
	
	{\bf \large C.~J. Halcrow$^*$ and J.~I. Rawlinson$^\dagger$} \\[1pt]
	\vskip 1em
	{\it 
		$^*$School of Mathematics, University of Leeds, U.K. \linebreak email address: c.j.halcrow@leeds.ac.uk} \\
	\vskip 1em
	{\it 
		$^\dagger$DAMTP,
		University of Cambridge, U.K. \linebreak email address:  jir25@damtp.cam.ac.uk} \\[20pt]
	
\end{center}

\abstract{}

We develop a formalism to calculate electromagnetic (EM) transition rates for rotational-vibrational models of nuclei. The formalism is applied to recently proposed models of Carbon-12 and Oxygen-16 which are inspired by nuclear dynamics in the Skyrme model. We compare the results to experimental data, as well as other nuclear models. The results for Carbon-12 are in good agreement with the data across all models, making it difficult to differentiate the models. More experimental data is needed to do this, and we suggest which transitions would be most interesting to measure. The models of Oxygen-16 are less successful in describing the data, and we suggest some possible improvements to our approximations which may help.

\vskip 5pt

\setcounter{page}{1}
\renewcommand{\thefootnote}{\arabic{footnote}}

\section{Introduction}

Understanding the intrinsic structure of nuclei is one of the central problems in nuclear physics. There is still much debate about the nature of light nuclei, even for stable abundant nuclei such as Carbon-12 and Oxygen-16. These are often described using $\alpha$-particle models \cite{Wheel}. Here, nucleons cluster into groups of four ($\alpha$-particles) and the nuclei have the symmetry of a simple geometric shape -- the $\alpha$-particles lying on the shape's vertices. Carbon-12 and Oxygen-16 are described as a triangle and tetrahedron respectively. The triangular model includes a low lying rotational band with spins $0^+,2^+,3^-,4^\pm,...$ for Carbon-12 while the tetrahedral model has one with spins $0^+,3^-,4^+,...$ for Oxygen-16. Both are seen experimentally, confirmed after the recent clarification of a $4^-$ state at 11.83 MeV \cite{Freer} and a $5^-$ state at 22.4 MeV \cite{Freer2} for Carbon-12. There is much debate about the higher energy states. For example, Carbon-12 has an approximate  higher energy rotational band with spins $0^+,2^+,4^+,...$ . Different authors model this band as a chain of $\alpha$-particles \cite{LM}, a ``breathing" excitation of the triangle \cite{acm12}, or an admixture of several shapes \cite{rawlinson}. All these models can reproduce the energy spectrum rather well.

Rotational bands are not the only indicator of collective, geometric behaviour. Electromagnetic (EM) transition rates measure $\gamma$-decay between two nuclear states. Here, the higher energy state emits a photon which carries away spin and energy. These decays are only seen below (or nearby) the strong decay threshold as they are electromagnetic in nature. Above this threshold, strong interactions dominate the decay paths. Theoretically the EM rates depend on the overlap of wavefunctions and the charge density multipole tensor. Generically, a large transition rate indicates collective behaviour. In fact, the large E3 transition rate between the low lying $3^-$ and $0^+$ states of Oxygen-16 is a motivation for the continuing interest in $\alpha$-particle models \cite{Rob}. Its size is unexplained in the basic shell model, where the decay strength should be close to a single Weisskopf unit, and in basic collective models, where the nucleus is described as a vibrating bag of nuclear matter \cite{Berg}.

Just as the EM transitions can help differentiate collective behaviour from single-particle behaviour, in this paper we will try and use them to differentiate between particular $\alpha$-particle models. Since the transition rates depend on the structure of the wavefunctions, physically different models should provide different results. To see these differences, we calculate the EM rates for recently proposed models of Carbon-12 \cite{rawlinson} and Oxygen-16 \cite{HKM}, which were inspired by nuclear dynamics in the Skyrme model. In these, sets of configurations are constructed which include several low lying shapes: the triangle and chain for Carbon-12 and the tetrahedron and square for Oxygen-16. The wavefunctions take values across the entire set of shapes, and can be interpreted physically as mixtures of the different geometric shapes.

The wavefunctions are rotational-vibrational states. The rotational symmetry of space manifests itself through rigid body wavefunctions and these are combined with vibrational wavefunctions, which account for deformations. We develop a formalism to calculate the transition rates for wavefunctions of this kind. The formalism applies to any model with an underlying ``shape" degree of freedom. The rigid body case, a common simplifying assumption in the Skyrme model \cite{ANW, HLM16} and $\alpha$-particle models \cite{BI2}, is a limiting case in our calculation. After developing this formalism in Section 2, we apply it to models of Carbon-12 and Oxygen-16 in Sections 3 and 4 respectively. These applications show the general nature of our work. The models are based on very different shape spaces: one is a $1$-dimensional graph made up of three edges joined at a single vertex while the other is a $2$-dimensional manifold. We compare our results to experimental data, as well as other nuclear models. Overall, each model gives very different results with different successes and failures when compared to data. We hope this theoretical work may motivate new experimental progress, as the latest data was taken in the early 1980s \cite{exp12,exp16}. We conclude with some further work and ideas in Section 5.

\section{General formalism}

We wish to describe nuclear dynamics by considering a large set of nuclear configurations with many possible shapes (the shape can be thought of as the nucleon distribution). We then choose a low energy subset of these configurations which we parametrise by a set of shape coordinates $\mathbf{s}$.
We also consider all possible orientations of these configurations
in physical space. Define coordinates as follows: for each shape,
choose a certain standard orientation of that shape in space (equivalently,
a body-fixed frame). Then parametrise all rotated versions of that
shape by Euler angles $\theta_{i}$ which specify the rotation that
relates the body-fixed frame to a space-fixed frame. In this fashion
we can define coordinates $\left(\mathbf{s},\theta_{i}\right)$.

Rotational symmetry of space means that quantum states can be classified by
a total angular momentum $J$ together with a space-fixed angular
momentum projection $J_{3}\in\left\{ -J,\ldots,+J\right\} $. States
$\ket{\Psi}$ within a given $\left(J,J_{3}\right)$ sector take the
form
\fontdimen16\textfont2=3.5pt
\fontdimen17\textfont2=3.5pt
\begin{equation}
\ket{\Psi}=\sum_{ L_{3}=-J}^{+J}\chi_{L_{3}}\left(\mathbf{s}\right)\ket{JJ_{3}L_{3}} \, , \label{eq:rovibstate}
\end{equation}
where we have expanded in a basis $\left\{ \ket{JJ_{3}L_{3}}\right\} $
of rigid-body wavefunctions which involve the body-fixed angular momentum
projection $L_{3}\in\left\{ -J,\ldots,+J\right\} $. These capture
the $\theta_{i}$ dependence of the state. The coefficient wavefunctions
$\chi_{L_{3}}\left(\mathbf{s}\right)$ satisfy a Schr\"odinger equation
defined on the space of shapes. We will see examples of this in the
specific models for Carbon-12 and Oxygen-16 considered in Sections
2 and 3.

\subsection{Electromagnetic transition rates}

In the long wavelength limit, the reduced transition probability for
electric multipole radiation between an initial state $\ket{i}$ of
spin $J$ and a final state $\ket{f}$ of spin $\tilde{J}$ is given
by \cite{greiner}

\begin{equation}
B\left(El,i\rightarrow f\right)=\frac{1}{2J+1}\sum_{J_{3},\tilde{J}_{3},m}\left|\int d^{3}r\bra{f}\rho\left(\mathbf{s},\mathbf{r},\theta_{i}\right)r^{l}Y_{lm}^{*}\left(\Omega\right)\ket{i}\right|^{2}\label{eq:bel}
\end{equation}
where $\mathbf{r}$ are space-fixed coordinates (with $\Omega$ the
angular coordinates in $\mathbf{r}$-space) and where $\rho\left(\mathbf{s},\mathbf{r},\theta_{i}\right)$
is the charge density of the configuration with shape $\mathbf{s}$
in orientation $\theta_{i}$. Note that the above expression involves
a sum over space-fixed spin projections $\tilde{J}_{3}$ for the final
state and an average over space-fixed spin projections $J_{3}$ for
the initial state.

We wish to calculate transition probabilities using (\ref{eq:bel})
for states of the form (\ref{eq:rovibstate}). The rigid-body wavefunctions
$\ket{JJ_{3}L_{3}}$ depend on Euler angles $\theta_{i}$ and so it will
help if we first simplify the $\theta_{i}$ dependence of the charge density
$\rho$. Expand $\rho$, evaluated at $\theta_{i}=\mathbf{0}$, in terms
of spherical harmonics
\begin{equation}
\rho\left(\mathbf{s},\mathbf{r},\mathbf{0}\right)=\sum_{l'=0}^{\infty}\sum_{m'=-l'}^{l'}c_{l'm'}\left(r\right)Y_{l'm'}\left(\Omega\right)
\end{equation}
where
\begin{equation}
c_{l'm'}\left(r\right)=\int d\Omega \, Y_{l'm'}^{*}\left(\Omega\right)\rho\left(\mathbf{s},\mathbf{r},\mathbf{0}\right).
\end{equation}
The spherical harmonics transform in a simple way under rotations,
giving the expression
\begin{equation}
\rho\left(\mathbf{s},\mathbf{r},\theta_{i}\right)=\sum_{l'}\sum_{m'}\sum_{m''}c_{l'm'}\left(r\right)Y_{l'm''}\left(\Omega\right)D_{m''m'}^{l'}\left(\theta_{i}\right)
\end{equation}
for the charge density in an arbitrary orientation $\theta_{i}$.
Substituting this into our original expression for $B\left(El,i\rightarrow f\right)$
gives 
\begin{equation}
B\left(El,i\rightarrow f\right)=\frac{1}{2J+1}\sum_{J_{3},\tilde{J}_{3},m}\left|\bra{f}\sum_{m'}D_{mm'}^{l}\left(\theta_{i}\right)\mathcal{Q}_{lm'}\left(\mathbf{s}\right)\ket{i}\right|^{2}
\end{equation}
where
\begin{equation} \label{qlmylm}
\mathcal{Q}_{lm}\left(\mathbf{s}\right)=\int d^{3}r\rho\left(\mathbf{s},\mathbf{r},\mathbf{0}\right)r^{l}Y_{lm}^{*}\left(\Omega\right) 
\end{equation}
is the multipole tensor of the charge density. This means that, for the initial state
\begin{equation} \label{initial}
\ket{i}=\sum_{L_{3}=-J}^{+J}\chi_{L_{3}}\left(\mathbf{s}\right)\ket{JJ_{3}L_{3}}
\end{equation}
and final state
\begin{equation}
\ket{f}=\sum_{\tilde{L_{3}}=-\tilde{J}}^{+\tilde{J}}\tilde{\chi}_{\tilde{L_{3}}}\left(\mathbf{s}\right)\ket{\tilde{J}\tilde{J_{3}}\tilde{L_{3}}},
\end{equation}
we have that
\begin{align}  
&B\left(El,i\rightarrow f\right) =  \frac{1}{2J+1}\sum_{J_{3},\tilde{J}_{3},m}\left|\sum_{m'}\bra{f}D_{mm'}^{l}\left(\theta_{i}\right)\mathcal{Q}_{lm'}\left(\mathbf{s}\right)\ket{i}\right|^{2} \nonumber \\
 &= \frac{1}{2J+1}\sum_{J_{3},\tilde{J}_{3},m}\left|\int d\mathbf{s}\sum_{m',L_{3},\tilde{L}_{3}}\tilde{\chi}_{\tilde{L_{3}}}^{*}\left(\mathbf{s}\right)\chi_{L_{3}}\left(\mathbf{s}\right)\mathcal{Q}_{lm'}\left(\mathbf{s}\right)\bra{\tilde{J}\tilde{J_{3}}\tilde{L_{3}}}D_{mm'}^{l}\left(\theta_{i}\right)\ket{JJ_{3}L_{3}}\right|^{2} \nonumber \\
 &= \frac{2\tilde{J}+1}{\left(2J+1\right)^{2}}\sum_{J_{3},\tilde{J}_{3},m}\left|\int d\mathbf{s}\sum_{m',L_{3},\tilde{L}_{3}}\tilde{\chi}_{\tilde{L_{3}}}^{*}\left(\mathbf{s}\right)\chi_{L_{3}}\left(\mathbf{s}\right)\mathcal{Q}_{lm'}\left(\mathbf{s}\right)\bigl\langle\tilde{J}\tilde{J}_{3}lm\bigr|JJ_{3}\bigr\rangle\bigl\langle\tilde{J}\tilde{L}_{3}lm'\bigr|JL_{3}\bigr\rangle\right|^{2}\nonumber \\
 &= \frac{2\tilde{J}+1}{2J+1}\left|\int d\mathbf{s}\sum_{m',L_{3},\tilde{L}_{3}}\tilde{\chi}_{\tilde{L_{3}}}^{*}\left(\mathbf{s}\right)\chi_{L_{3}}\left(\mathbf{s}\right)\mathcal{Q}_{lm'}\left(\mathbf{s}\right)\bigl\langle\tilde{J}\tilde{L}_{3}lm'\bigr|JL_{3}\bigr\rangle\right|^{2}  \label{Bform}
\end{align}
where $\bigl\langle\tilde{J}\tilde{J}_{3}lm\bigr|JJ_{3}\bigr\rangle$ are Clebsch-Gordan coefficients and in the final equality we used
\begin{equation}
 \sum_{J_{3},\tilde{J}_{3},m}\left|\bigl\langle\tilde{J}\tilde{J}_{3}lm\bigr|JJ_{3}\bigr\rangle\right|^{2}=2J+1
\end{equation}
whenever $J = \tilde{J} + l,\ldots,|\tilde{J}-l| $. For values of $J$ outside of this range, the Clebsch-Gordan coefficients all vanish and the transition rate is zero. We have now written the original expression in terms of an overlap between vibrational wavefunctions, weighted by the charge density multipole tensor and some Clebsch-Gordan coefficients. All these are relatively straightforward to calculate, even if the expression is rather complicated. Note that for $\tilde{J}=0$ the expression $(2.10)$ simplifies (using $\bigl\langle00lm'\bigr|JL_{3}\bigr\rangle=\delta_{Jl}\delta_{L_{3}m'}$)
to give
\begin{equation}
B\left(El,i\rightarrow f\right)=\frac{\delta_{Jl}}{2J+1}\left|\int d\mathbf{s}\,\tilde{\chi}_{0}^{*}\left(\mathbf{s}\right)\sum_{L_{3}}\chi_{L_{3}}\left(\mathbf{s}\right)\mathcal{Q}_{lL_{3}}\left(\mathbf{s}\right)\right|^{2} \, ,
\end{equation}
which mimics the structure of the initial wavefunction \eqref{initial}.

We also note here that
\begin{align}
&B\left(El,f\rightarrow i\right)  =  \frac{2J+1}{2\tilde{J}+1}\left|\int d\mathbf{s}\sum_{m',L_{3},\tilde{L}_{3}}\tilde{\chi}_{L_{3}}^{*}\left(\mathbf{s}\right)\chi_{\tilde{L}_{3}}\left(\mathbf{s}\right)\mathcal{Q}_{lm'}\left(\mathbf{s}\right)\bigl\langle JL_{3}lm'\bigr|\tilde{J}\tilde{L}_{3}\bigr\rangle\right|^{2} \nonumber \\
  &=  \left|\int d\mathbf{s}\sum_{m',L_{3},\tilde{L}_{3}}\tilde{\chi}_{L_{3}}^{*}\left(\mathbf{s}\right)\chi_{\tilde{L}_{3}}\left(\mathbf{s}\right)\mathcal{Q}_{lm'}\left(\mathbf{s}\right)\left(-1\right)^{m'}\bigl\langle\tilde{J}\tilde{L}_{3}l\left(-m'\right)\bigr|JL_{3}\bigr\rangle\right|^{2} \nonumber \\
  &=  \frac{2J+1}{2\tilde{J}+1}B\left(El,i\rightarrow f\right) \, ,
\end{align}
where we have used symmetry properties of the Clebsch-Gordan coefficients
together with the identity $Y_{lm}^{*}\left(\Omega\right)=(-1)^{m}Y_{l\left(-m\right)}\left(\Omega\right)$.

\subsection{Estimating $\mathcal{Q}$ for point $\alpha$-particle models}

The nuclear models we will consider in Sections 3 and 4 are based on configurations of $\alpha$-particles. For the purposes of calculating electromagnetic transition rates, we will treat these $\alpha$-particles as point charges. For $\alpha$-particles at positions $\mathbf{R}_{1}\left(\mathbf{s}\right),\ldots,\mathbf{R}_{N}\left(\mathbf{s}\right)$,
we therefore approximate the charge density by 
\begin{equation}
\rho\left(\mathbf{s},\mathbf{r},\mathbf{0}\right)=\sum_{i=1}^{N}2\delta^{\left(3\right)}\left(\mathbf{R}_{i}\left(\mathbf{s}\right)-\mathbf{r}\right).
\end{equation}
Substituting this into \eqref{qlmylm} leads to the multipole tensor
\begin{equation} \label{Qlm}
\mathcal{Q}_{lm}\left(\mathbf{s}\right)=\sum_{i=1}^{N}2R_{i}\left(\mathbf{s}\right)^{l}Y_{lm}^{*}\left(\hat{\mathbf{R}}_{i}\left(\mathbf{s}\right)\right).
\end{equation}

\section{Quantum graph model for Carbon-12}

\subsection{Introduction}

Theoretical studies of the Carbon-12 nucleus have a long and interesting
history. Most famously, in the 1950's Fred Hoyle predicted that Carbon-12
should have a positive-parity resonance just above the threshold for
breakup into Beryllium-8 and Helium-4. He argued that such a state
would lead to resonant enhancement of Carbon-12 production during
stellar nuclear synthesis, explaining the abundance of Carbon-12 in
our universe. His prediction was confirmed experimentally with the
discovery of the $7.7$ MeV $0^{+}$ excitation, now known as the
Hoyle state. 

It is widely agreed that Carbon-12 can be usefully thought
of in terms of alpha clusters. There is a band in the observed energy spectrum containing states with the characteristic spin and parity combinations
$0^{+}$, $2^{+}$, $3^{-}$, $4^\pm,\ldots$ often referred to as the ground state band. These are exactly the states which arise from a rotating equilateral triangle of $\alpha$-particles, and are physically interpreted as such. There has been less agreement on the physical interpretation of the Hoyle state (and the other observed low-lying
excited states outside of the ground state band) with many interpretations
offered including a rigid linear chain \cite{LM} , a bent-arm \cite{abin}, a breathing vibration 
of an equilateral triangle \cite{acm12} and even a diffuse gas of $\alpha$-particles \cite{gas}.
All can give a reasonable fit to the observed energy spectrum of Carbon-12
and so electromagnetic transition strengths are our best hope for distinguishing
these models.

The quantum graph model (QGM) for Carbon-12, introduced in \cite{rawlinson},
is based on the quantized dynamics of three point $\alpha$-particles.
The QGM allows for isosceles triangles of $\alpha$-particles
which interpolate between the equilateral triangle and linear chain clusters
and so includes both of these highly symmetric configurations along
with the intermediate bent-arm (obtuse triangle) configurations. There
are three ways in which an equilateral triangle cluster of $\alpha$-particles
can be deformed into a chain, because any one of the three $\alpha$-particles
can become the middle $\alpha$-particle in the chain. Thus the space
of allowed shapes corresponds to a three-edged graph as shown in Figure 1. 

\begin{figure}
\begin{centering}
\includegraphics[scale=0.8]{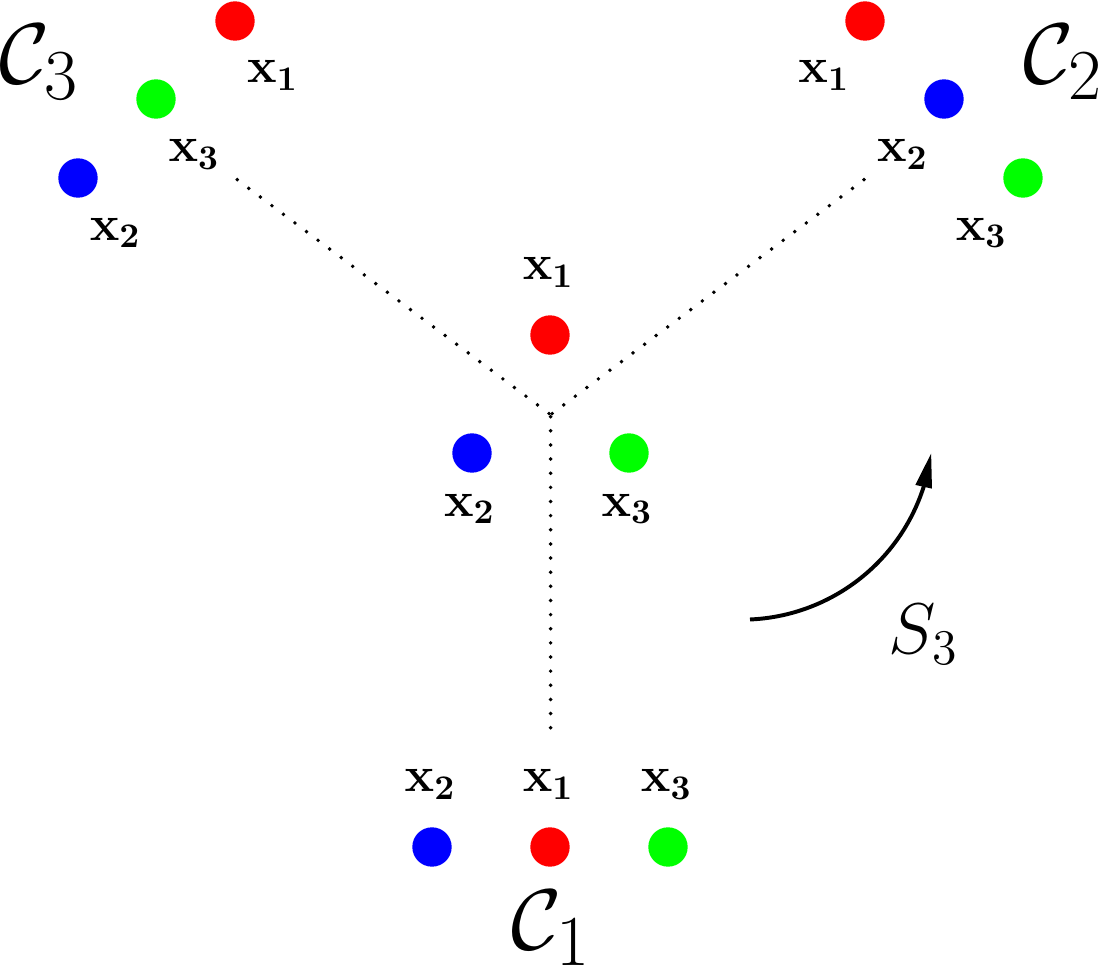}
\par\end{centering}
\caption{The graph of configurations for the QGM of Carbon-12. The central configuration is an equilateral triangle. This interpolates into three different chain configurations along the three graph edges. }
\label{graphillustration}
\end{figure}

In more detail, the space of shapes is defined as follows: we restrict
configurations of three point $\alpha$-particles to those isosceles triangles
which interpolate between an equilateral triangle and a linear chain.
The equilateral triangle corresponds to the vertex of the graph. The
equilateral triangle can deform in three ways, corresponding to the
three edges leaving the vertex. Focusing on a particular edge (labelled
$\mathcal{C}_{1}$ in Figure \ref{graphillustration}), we define
a shape coordinate $s$ on this edge such that the positions $\mathbf{R}_{i}\left(s\right)$
of the three $\alpha$-particles are
\begin{align}
\mathbf{R}_{1}&=f\left(s\right)\left(0,s,0\right) \\ 
\mathbf{R}_{2}&=f\left(s\right)\left(-\frac{1}{2}\sqrt{2-3s^{2}},-\frac{1}{2}s,0\right) \\
\mathbf{R}_{3}&=f\left(s\right)\left(\frac{1}{2}\sqrt{2-3s^{2}},-\frac{1}{2}s,0\right).
\end{align}
The $\mathbf{R}_{i}$ determine the standard orientation at the point $s$ on the graph. Here $f\left(s\right)\approx 1.1 - 0.2s$ is a linear function of $s$ which fixes the
overall scale of the triangle relative to the linear chain, as discussed in \cite{rawlinson}. The range
we consider is $s\in\left[0,s_{\mathrm{max}}\right]$ where $s_{\mathrm{max}}=\frac{1}{\sqrt{3}}.$
Note that $s=0$ gives a linear chain cluster and as we increase $s$
we approach an equilateral triangle cluster at $s=s_{\mathrm{max}}$.
By acting on these configurations with rotations, we can generate
all possible orientations of these shapes . We use coordinates
$\left(s,\theta_{i}\right)$ with Euler angles $\theta_{i}$ describing
the rotation relating a given configuration to these standard configurations.
A similar construction is carried out on the other two edges, and
the union of all three of these gives the total configuration space
$\mathcal{C}$. The three $\alpha$-particles should be indistinguishable:
this is imposed at the quantum level by demanding that states lie in the trivial representation of the group $S_{3}$ which
acts on $\mathcal{C}$ by permuting the three particles. Quantization
requires ideas from Quantum Graph Theory, as explained in \cite{rawlinson}. Briefly, the wavefunction on edge $\mathcal{C}_{1}$ can be expanded in terms of rigid body states as
\begin{equation}
\ket{\Psi}=\sum_{L_{3}=-J}^{+J}\chi_{L_{3}}\left(\mathbf{s}\right)\ket{JJ_{3}L_{3}}
\end{equation} where the $\chi_{L_{3}}$ satisfy a Schr\"odinger equation, and Quantum Graph Theory boundary conditions are imposed at the vertex.

\begin{figure}[h!]
	\includegraphics[width=0.95\textwidth]{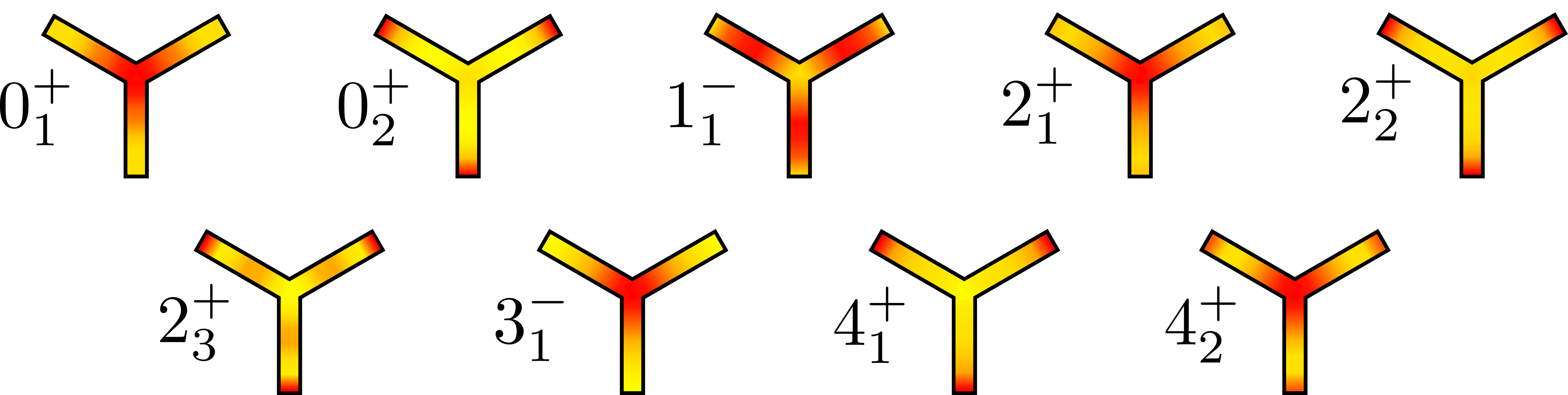}
	\begin{centering}
		\caption{Shape probability densities. The colours red and yellow correspond to regions of high and low probability density. Each density is rescaled so that the maximum of the wavefunction is red. Hence, for example, the $0_2^+$ state is highly concentrated while the $1_1^-$ state is more evenly spread.  }
		\label{probd}
		\par\end{centering}
		
\end{figure}

Permutation symmetry restricts the form of the wavefunctions on the
edge $\mathcal{C}_{1}$. The allowed states, relevant for our
calculation, are listed in Table \ref{wavefncs12}. For each state we calculate a shape probability density, defined as
\begin{equation}
P_\Psi(\boldsymbol{s}) = \sum_{L_3=-J}^{J} |\chi_{L_3}(\boldsymbol{s})|^2 \, .
\end{equation}
We plot the shape probability density function for each of the wavefunctions in Figure \ref{probd}. The physical interpretation of states can be seen by looking at which shapes these are concentrated at. For example, the $0_1^+$ state is interpreted as an equilateral triangular state while the $0_2^+$ state is concentrated at the linear chain. The $1_1^-$ state is forbidden at both of these shapes and is instead concentrated at an intermediate bent-arm configuration.

\begin{table}
\begin{centering}
\begin{tabular}{|c|c|c|}
\hline
$J^{P}$ & Wavefunction & $E_{\mathrm{exp}}$(MeV) \tabularnewline
\hline 
$0_{1}^{+}$ & $\chi_{0}^{\left(0_{1}\right)}\left(s\right)\ket{0,0}$ & $0$ \tabularnewline
$0_{2}^{+}$ & $\chi_{0}^{\left(0_{2}\right)}\left(s\right)\ket{0,0}$ & $7.7$ \tabularnewline
$1_{1}^{-}$ & $\chi_{1}^{\left(1_{1}\right)}\left(s\right)\left(\ket{1,1}+\ket{1,-1}\right)$ & $10.8$ \tabularnewline
$2_{1}^{+}$ & $\chi_{2}^{\left(2_{1}\right)}\left(s\right)\left(\ket{2,2}+\ket{2,-2}\right)+\chi_{0}^{\left(2_{1}\right)}\left(s\right)\ket{2,0}$ & $4.4$ \tabularnewline
$2_{2}^{+}$ & $\chi_{2}^{\left(2_{2}\right)}\left(s\right)\left(\ket{2,2}+\ket{2,-2}\right)+\chi_{0}^{\left(2_{2}\right)}\left(s\right)\ket{2,0}$ & $9.9$ \tabularnewline
$2_{3}^{+}$ & $\chi_{2}^{\left(2_{3}\right)}\left(s\right)\left(\ket{2,2}+\ket{2,-2}\right)+\chi_{0}^{\left(2_{3}\right)}\left(s\right)\ket{2,0}$ & $16.1$ \tabularnewline
$3_{1}^{-}$ & $\chi_{3}^{\left(3_{1}\right)}\left(s\right)\left(\ket{3,3}+\ket{3,-3}\right)+\chi_{1}^{\left(3_{1}\right)}\left(s\right)\left(\ket{3,1}+\ket{3,-1}\right)$ & $9.6$ \tabularnewline
$4_{1}^{+}$ & $\chi_{4}^{\left(4_{1}\right)}\left(s\right)\left(\ket{4,4}+\ket{4,-4}\right)+\chi_{2}^{\left(4_{1}\right)}\left(s\right)\left(\ket{4,2}+\ket{4,-2}\right)+\chi_{0}^{\left(4_{1}\right)}\left(s\right)\ket{4,0}$ & $13.3$ \tabularnewline
$4_{2}^{+}$ & $\chi_{4}^{\left(4_{2}\right)}\left(s\right)\left(\ket{4,4}+\ket{4,-4}\right)+\chi_{2}^{\left(4_{2}\right)}\left(s\right)\left(\ket{4,2}+\ket{4,-2}\right)+\chi_{0}^{\left(4_{2}\right)}\left(s\right)\ket{4,0}$ & $14.1$ \tabularnewline
\hline
\end{tabular}
\par\end{centering}
\caption{The wavefunctions, in terms of vibrational wavefunctions and spin states, for each of the states considered in this paper. Each model state is identified with an experimental state, whose energy is also tabulated.  We suppress the $J_3$ label for ease of reading.}
\label{wavefncs12}

\end{table}

\subsection{Calculating $B\left(El\right)$ transition rates}

As an example, suppose we are interested in calculating $B\left(E3,3_{1}^{-}\rightarrow0_{1}^{+}\right)$
where $3_{1}^{-}$ denotes the lowest energy $J^{P}=3^{-}$ state and $0_{1}^{+}$
denotes the lowest $J^{P}=0^{+}$ state. The initial and final state
wavefunctions are
\begin{equation}
\ket{3_{1}^{-}}=\chi_{3}^{\left(3_{1}\right)}\left(s\right)\left(\ket{3J_{3}3}+\ket{3J_{3}-1}\right)+\chi_{1}^{\left(3_{1}\right)}\left(s\right)\left(\ket{3J_{3}1}+\ket{3J_{3}-1}\right)
\end{equation}
and
\begin{equation}
\ket{0_{1}^{+}}=\chi_{0}^{\left(0_{1}\right)}\left(s\right)\ket{000}.
\end{equation}
The expression $(2.10)$ from Section 2 gives
\begin{align*}
B\left(E3,3_{1}^{-}\rightarrow0_{1}^{+}\right)  =  \frac{1}{7}\Bigg|\int ds\, &{\chi_{0}^{\left(0_{1}\right)}}^{*}\left(s\right)\chi_{3}^{\left(3_{1}\right)}\left(s\right)\left(\mathcal{Q}_{33}\left(s\right)+\mathcal{Q}_{3-3}\left(s\right)\right)  \\ &+ {\chi_{0}^{\left(0_{1}\right)}}^{*}\left(s\right)\chi_{1}^{\left(3_{1}\right)}\left(s\right)\left(\mathcal{Q}_{31}\left(s\right)+\mathcal{Q}_{3-1}\left(s\right)\right)\Bigg|^{2}.
\end{align*}
In order to evaluate this integral we use the analytic expression for
\begin{equation*}
\mathcal{Q}_{lm}\left(s\right)=\sum_{i=1}^{3}2R_{i}\left(s\right)^{l}Y_{lm}^{*}\left(\hat{\mathbf{R}}_{i}\left(s\right)\right) \, ,
\end{equation*} treating the $\alpha$-particles as point particles as described in Section 2. The integration against the numerically generated wavefunctions $\chi_{L_{3}}\left(s\right)$ can be done over a single edge of the graph due to the symmetry of the system.

\subsection{Results}

The electromagnetic transition rates for the QGM are
displayed in Table \ref{carbon12results}. We pick the conversion
factor between fm and the length units in our model to be $\kappa=\sqrt{10}$.
Our results are displayed alongside results from an ab initio calculation
\cite{epelbaum} and the Algebraic Cluster Model (ACM) \cite{acm12},
along with a comparison to available experimental data. The ACM makes use of a bosonic quantization approach to the many-body problem.
It is based on an equilibrium configuration of $\alpha$-particles at
the vertices of an equilateral triangle, although allowing for large
rotation-vibration effects. The ab initio results are from Monte Carlo
lattice calculations based on chiral effective field theory. The authors
only consider four states: $0_{1}^{+}$, $2_{1}^{+}$, $0_{2}^{+}$
and $2_{2}^{+}$. The $0_{1}^{+}$ and $2_{1}^{+}$
states have a large overlap with a compact triangular arrangement
of $\alpha$-particles, so are interpreted physically as triangular states. In particular, the $2_1^+$ is interpreted as a rotational excitation of the $0_1^+$ state. The $0_{2}^{+}$ and $2_{2}^{+}$ states
have a large overlap with a bent-arm configuration (an obtuse triangle)
of $\alpha$-particles and are interpreted as the first two states on a rotational band of this shape. This is consistent with the results of the QGM.

\begin{table}
	\begin{centering}
		\begin{tabular}{|c|c|c|c|c|}
			\hline 
			$B\left(El,i\rightarrow f\right)$ & QGM & ab  & ACM \cite{acm12} & experiment \tabularnewline
			& $\left(\kappa=\sqrt{10}\right)$ &  initio \cite{epelbaum}  &   &  $\left[e^{2}\text{fm}^{2l}\right]$\cite{exp12}   \tabularnewline
			\hline 
			$B\left(E2,2_{1}^{+}\rightarrow0_{1}^{+}\right)$  & $11.7$ & $5$ & $8.4$ & $7.6\pm0.42$\tabularnewline
			$B\left(E3,3_{1}^{-}\rightarrow0_{1}^{+}\right)$  & $62.4$ & & $44$ & $103\pm13.7$\tabularnewline
			$B\left(E4,4_{1}^{+}\rightarrow0_{1}^{+}\right)$  & $170$ &  & $73$ & \tabularnewline
			$B\left(E2,2_{2}^{+}\rightarrow0_{1}^{+}\right)$  & $1.16$ & $2$ &  & \tabularnewline
			$B\left(E4,4_{2}^{+}\rightarrow0_{1}^{+}\right)$  & $11.6$ &  &  & \tabularnewline
			$B\left(E2,2_{3}^{+}\rightarrow0_{1}^{+}\right)$  & $0.408$ &  &  & $0.67\pm0.13$\tabularnewline
			$B\left(E2,2_{1}^{+}\rightarrow0_{2}^{+}\right)$  & $1.10$ & $1.5$ & $0.26$ & $2.7\pm0.28$\tabularnewline
			$B\left(E2,2_{2}^{+}\rightarrow0_{2}^{+}\right)$ & $24.7$ & $6$ &  & \tabularnewline
			$B\left(E1,2_{3}^{+}\rightarrow1_{1}^{-}\right)$  & $0$ &  &  & $\left(3.1\pm0.78\right)\times10^{-3}$\tabularnewline
			$B\left(E1,2_{3}^{+}\rightarrow3_{1}^{-}\right)$ & $0$ &  &  & $\left(1.1\pm0.20\right)\times10^{-3}$\tabularnewline
			$B\left(E1,2_{1}^{+}\rightarrow3_{1}^{-}\right)$ & $0$ &  &  &  \tabularnewline
			
			\hline 
		\end{tabular}
		\par\end{centering}
	\caption{EM transition rates $B\left(El,i\rightarrow f\right)$ for Carbon-12. We tabulate the results for the model described in this Section, the ab initio calculation and the Algebraic Cluster Model, as well as the available experimental data. All values are in units of $e^2$fm$^{2l}$.}
	\label{carbon12results}
	
\end{table}

The structure of the transition rate formula \eqref{Bform} shows that the strength of the transition rate depends on the overlap between wavefunctions, as well as the multipole moments and structure of the wavefunctions. However, the final result of the calculation is difficult to predict before doing  it in full. For instance, the $0_1^+$ and $4_1^+$ states appear to have little overlap, as we can see in Figure \ref{probd}.  Due to this we might expect that the E4 transition, which links these states, would be small. However, the vibrational wavefunctions $\chi_{L_{3}} $ for both states have no nodes and so their product has the same sign at all points in configuration space. Hence the integrand doesn't change sign anywhere and this constructive interference between wavefunctions leads to a large integral. In contrast, the $0_1^+$ and $4_2^+$ states appear to have a large overlap. However, the $4_2^+$ vibrational wavefunctions change sign. This leads to an integrand with both positive and negative parts which interference destructively, giving the small result.

Along the ground state band ($0_{1}^{+}$, $2_{1}^{+}$, $3_{1}^{-}$,
...) there is no major discrepancy between the various models. The agreement is expected as all the models have a similar interpretation of the ground state band as arising from a rotating equilateral triangle. The results along the ground state band are also in broad agreement with experimental data, although all models slightly underestimate the $E3$ transition.

The $B\left(E1,2^{+}\rightarrow3^{-}\right)$ and $B\left(E1,2^{+}\rightarrow1^{-}\right)$ transition
strengths come out as zero in our model due to the symmetries of the wavefunctions. This is also true for the ACM and simple geometric models, as shown in \cite{SFV16}. The authors study the representation theory underlying transition rate calculations - giving selection rules and in particular rules out $E1$ transitions for Carbon-12. This is consistent with the very small observed
values $\sim10^{-3}$ $e^{2}$fm$^{2}$. For the states that have been experimentally measured, there is little to distinguish the models. Because of this, we must instead look at transitions for states that have not yet been measured. The $B(E2; 2_2^+ \to 0_2^+)$ transition is four times larger for us compared to the ab initio prediction. We expect the transition will also be smaller in the ACM. This transition is therefore a key data point which would distinguish the various models.

The most significant difference between experiment and theory is seen for \linebreak $B\left(E2,2_{1}^{+}\rightarrow0_{2}^{+}\right)$, the transition
between the Hoyle state and the ground state band. Here the ACM value
is too small by a factor of $10$. Our model and the ab initio
calculation do better than the ACM here, although we still underestimate
the value slightly. Recall that the ab initio approach finds a large
overlap of the Hoyle state with an obtuse triangular configuration.
Our work supports this interpretation, with the $0_{2}^{+}$ wavefunction
peaking at the linear chain but allowing a superposition of shapes
near to the chain. The picture in the ACM is different, with the Hoyle
state interpreted as a breathing excitation of the equilateral triangle.
More data is needed, both experimental and from competing models,
in order to make further comparisons and we hope that our calculations
will stimulate further work in this direction.

\section{E-manifold model for Oxygen-16}

\begin{figure}[h!]
	\begin{centering}
		\includegraphics[width=0.9\textwidth]{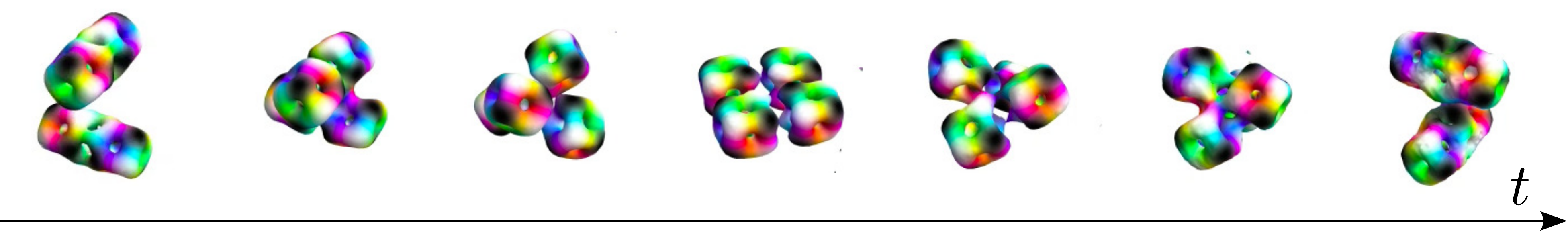}
		\par\end{centering}
	\caption{A numerically generated scattering path which links asymptotic configurations to the tetrahedron, the flat square and the dual tetrahedron. The dynamics are generated from the Skyrme model and we plot contours of the energy density. Time evolution is read left to right.}
	\label{Scattering}
\end{figure}

Since Wheeler's pioneering work, Oxygen-16 has often been modeled as a tetrahedron of $\alpha$-particles \cite{Wheel}. Later, sophisticated $\alpha$-models found that other low energy geometric configurations exist, including the $4\alpha$-chain, the flat square and the bent square \cite{Bau}. In fact, the final two are closely related to the tetrahedron. All these are joined by a dynamical mode, shown in Figure \ref{Scattering}. We'll now review a model, first constructed in \cite{HKM}, which accounts for the configurations which appear in this Figure. In fact, this path is part of a two-dimensional manifold which we'll call the E-manifold. The manifold can be visualised as a sphere with 6 punctures, and we model it as the 6-punctured sphere with negative constant curvature. The position on the punctured sphere $(x,y,z)$ corresponds to the position of one of the $\alpha$-particles. The other three then lie at $(x,-y,-z), (-x,y,-z)$ and $(-x,-y,z)$. This fixes the standard orientation of the configurations. For instance, the point $(x,y,z)=(1,1,1)$ corresponds to a tetrahedron, while the point $(x,y,z) = (1,1,0)$ represents a flat square.

\begin{figure}[h!]
	\centering
	$\begin{array}{l}
	\includegraphics[scale=0.5,keepaspectratio=true]{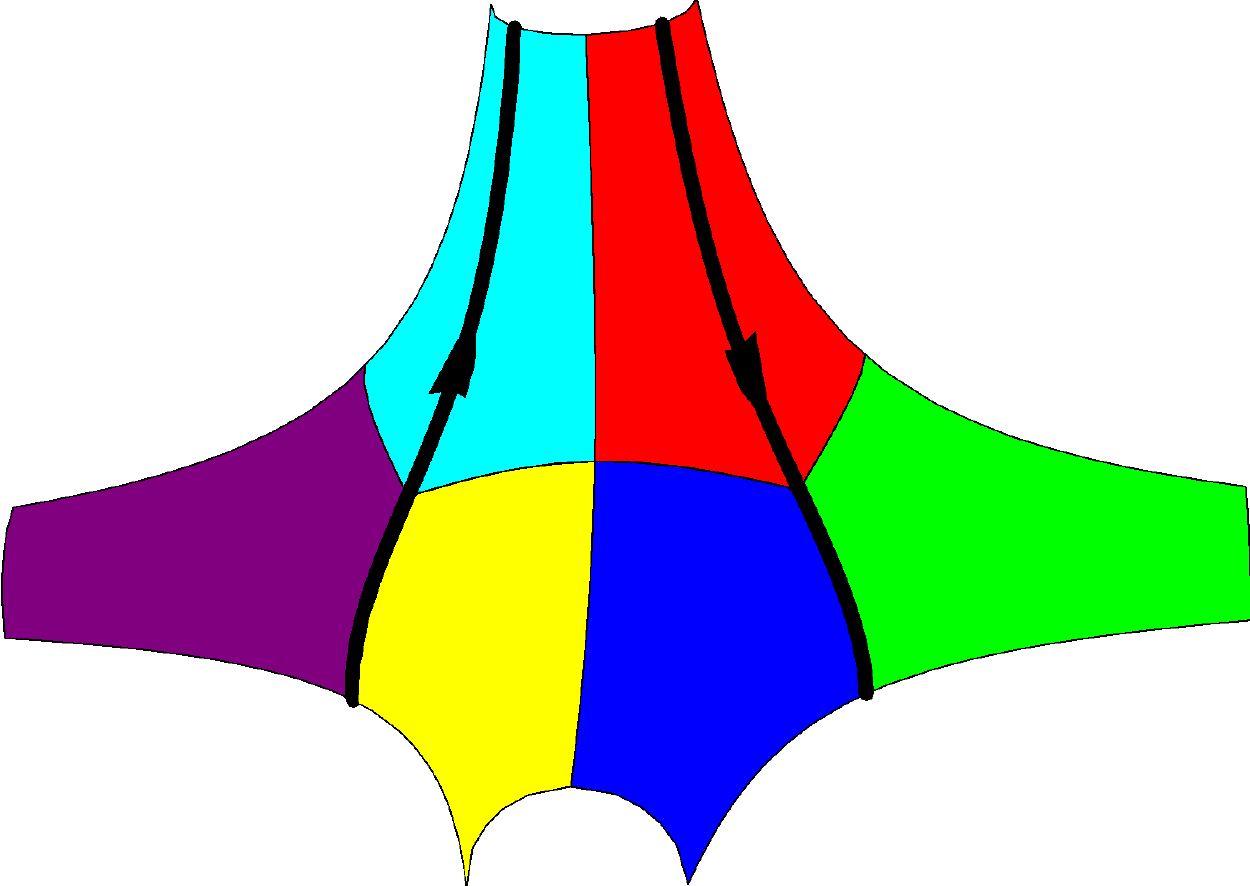}
	\label{fig:equiv}
	\end{array}$
	$\qquad \cong \qquad$
	$\begin{array}{l}
	\includegraphics[scale=0.35,keepaspectratio=true]{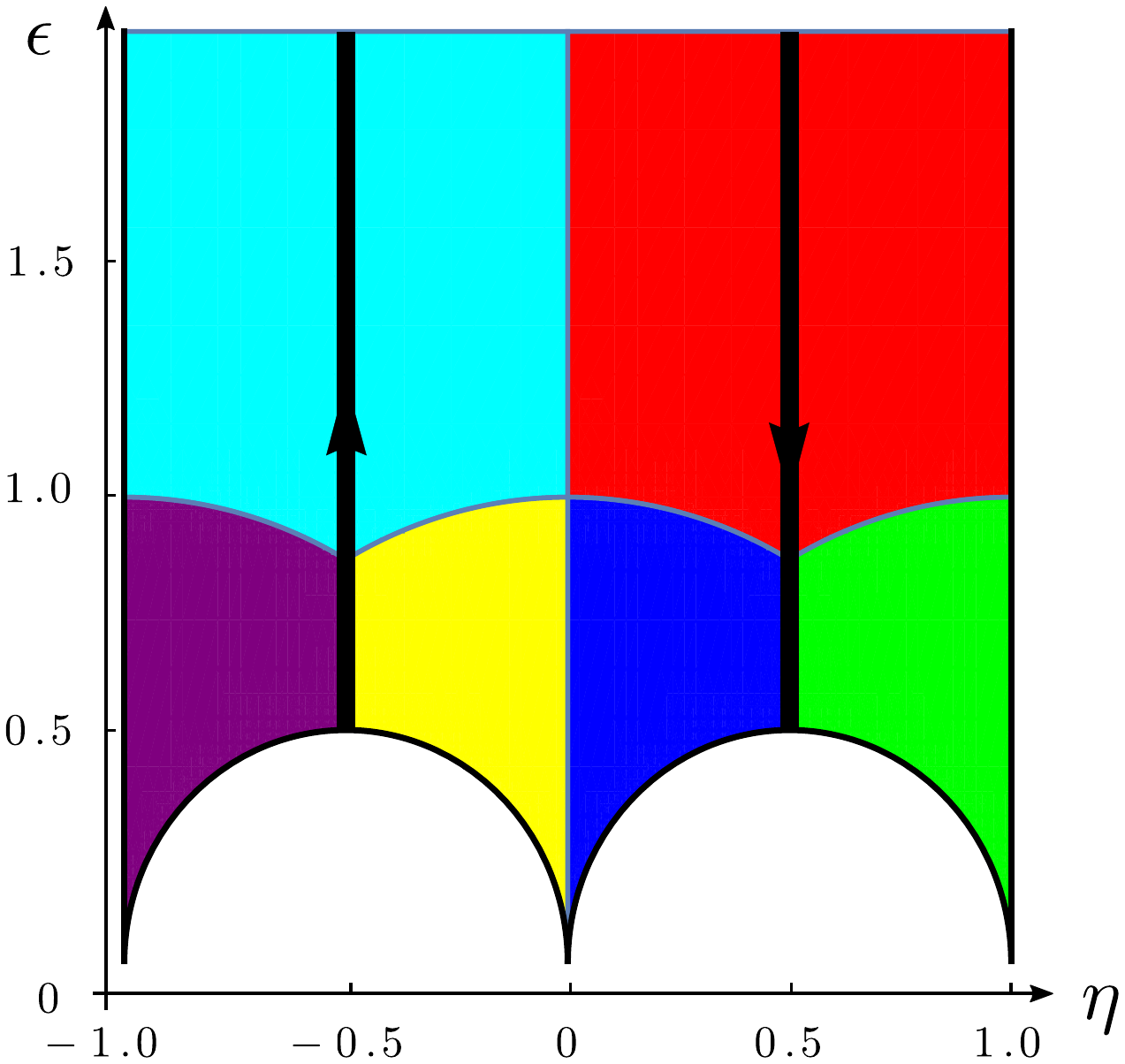}
	\end{array}$
	\caption{The relation between a quarter of the six-punctured sphere (left) and a portion of the complex plane (right). Tetrahedral configurations are at the points where three coloured regions meet while the square configurations are at points where four coloured regions meet. The scattering mode from Fig. \ref{Scattering} is represented by the thick black lines.}
	\label{equiv}
\end{figure}

Since knowing one particle's position automatically fixes the other three, we can focus on one quarter of the sphere. Using hyperbolic geometry we can then project this quarter sphere onto a portion of the complex plane. This mapping is displayed in Figure \ref{equiv}, where the positions of the geometric shapes, as well as the dynamical path from Figure \ref{Scattering}, are also plotted. We will use $\zeta = \eta + i\epsilon$ as the coordinates on the complex plane. 

The EM transition rates depend on the wavefunction and the multipole moments of the charge density, $\mathcal{Q}_{lm}$. Hence we must write these in terms of $\eta$ and $\epsilon$. As explained in Section 2.2, we can write $\mathcal{Q}_{lm}$ in terms of the positions of the particles, so we must find the mapping between the particle positions and the complex variables. We do this now. Given a point $\zeta$ on the complex plane, the position on a unit sphere is given by
\begin{equation}
(X,Y,Z) = \frac{1}{1+\left| H(\zeta) \right|^{2}}\left( 2\, \text{Re}\left(H(\zeta)\right) , 2\,\text{Im}(H\left(\zeta)\right),1 - \left| H(\zeta) \right|^{2}  \right) \, ,
\end{equation}
where
\begin{equation}
H(\zeta) = \left(\frac{\Theta_3 \left(   \pi / 4 , \exp\left( i \pi \zeta \right) \right)}{  \exp\left( \pi i (1+\zeta)/4 \right) \Theta_3\left( \pi(1+2\zeta)/4, \exp(i \pi \zeta) \right)  }  \right)^2 \, ,  
\end{equation}
and $\Theta_3$ is a Jacobi theta function \cite{FK01}. Having found the positions on a unit sphere, these should now be projected onto a sphere with punctures. We have some choice in this map but are constrained physically. We know the moments of inertia of the tetrahedron and square within the Skyrme model \cite{CJH}. Additionally, once the configuration breaks into two clusters (as in the far left and far right of Figure \ref{Scattering}) one of the moments must become constant and the other two grow quadratically with distance. The following projection satisfies all the aforementioned conditions 
\begin{equation}
\mathbf{R}_1 = \frac{\kappa}{\sqrt{1-(\text{max}(X,Y,Z))^2}}\left( X,Y,Z \right)  \, .
\end{equation}
The constant $\kappa$ gives the scale of the configuration. As an example, to calculate the positions of the $\alpha$-particles at $\zeta = 0 + i$, we first calculate $H(i) = 1+\sqrt{2}$, giving a unit sphere coordinate $(2^{-1/2},0,-2^{-1/2} )$. We then map this to the position $\mathbf{R}_1 = \kappa(1,0,-1)$. This is the position of one of the particles; the other three lie at $\boldsymbol{R}_2=\kappa(1,0,1),\boldsymbol{R}_3= \kappa(-1,0,-1)$ and $\boldsymbol{R}_4=\kappa(-1,0,1)$. Hence the point $\zeta=i$ corresponds to a flat square, lying in the $x$-$z$ plane. We use these values of $\mathbf{R}_i$ to calculate $\mathcal{Q}_{lm}(\zeta)$ using equation \eqref{Qlm}. The scale parameter $\kappa$ is later fixed, to match the $B(E3; 3_1^- \to 0_1^+)$ transition rate.

\begin{figure}[h!]
		\includegraphics[width=1.0\textwidth]{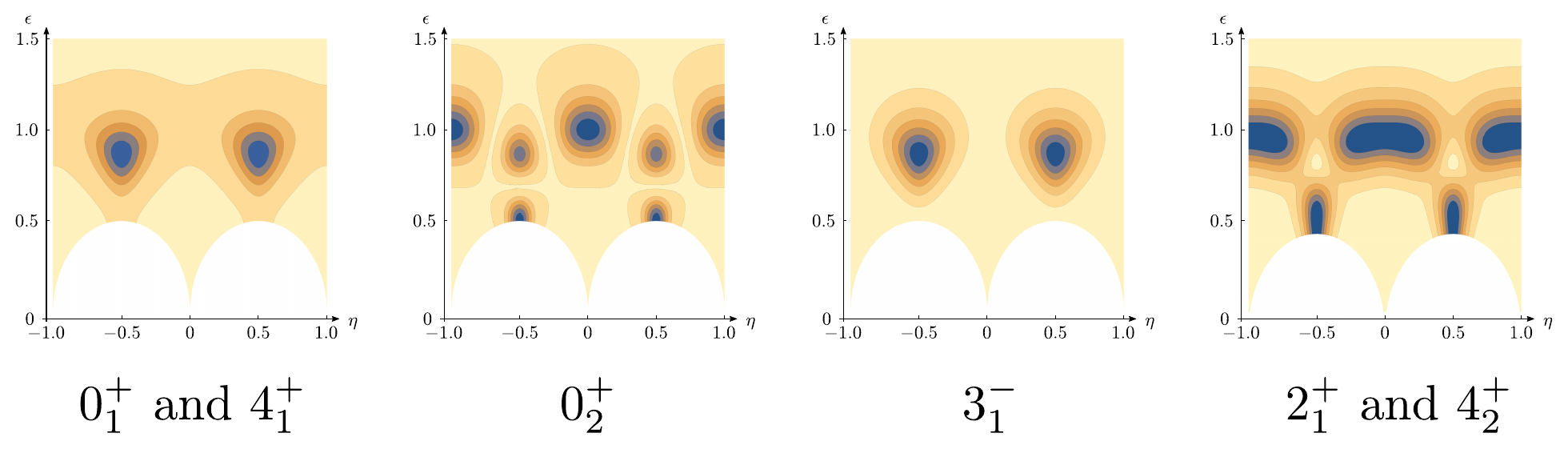}

	\caption{Shape probability densities for each wavefunction, plotted on a region of the complex $\zeta$-plane. Blue regions correspond to large densities while pale regions have small densities.}
	\label{probdent}
\end{figure}

To find quantum states we must first calculate vibrational wavefunctions on the complex plane. These satisfy a Schr\"odinger equation which in turn depends on a metric and potential on the E-manifold of configurations. These were fixed in \cite{HKM} and the Schr\"odinger equation takes the form
\begin{equation} \label{Schro}
-\frac{\hbar^2}{2}\epsilon^2\left( \frac{\partial^2}{\partial \eta^2} + \frac{\partial^2}{\partial\epsilon^2}\right)\psi + \epsilon^2\left(\frac{1}{2}\omega^2\left(\eta-\frac{1}{2}\right)^2 +\mu^2 \right)\psi = E_\text{vib} \psi \, ,
\end{equation}
where $\omega$ and $\mu$ are phenomenological parameters. The potential was chosen so that the tetrahedra have minimal energy, the squares have higher energy (by around $6$ MeV) and the asymptotic configurations have even higher energy. The expression \eqref{Schro} is only valid in the red region of the complex plane (for the colouring, see Figure \ref{equiv}). The wavefunctions were calculated in \cite{HKM} and classified further in \cite{HKM2}. Four of them will be relevant for our calculation - labeled $\psi_{T0}^+, \psi_{T1}^+, \psi_{S0}^-$ and $(u_1^+,v_1^+)$. These are combined with rigid-body wavefunctions to create physical states. The allowed states, relevant for our calculation, are listed in Table 3. We plot the shape probability density function on the complex plane for each of the wavefunctions in Figure \ref{probdent}. We sometimes say that a state is ``tetrahedral" or ``square-like". This means that the corresponding probability density is concentrated at those configurations. The states $0_1^+,3_1^-,4_1^+$ are all tetrahedral and form an approximate rotational band. The states $2_1^+$ and $4_1^+$ are both strongly concentrated at the squares and should be thought of as rotational excitations of a square configuration. The $0_2^+$ state is concentrated at squares and tetrahedra, and is interpreted as an admixture of both these geometries.

\begin{table}
\begin{center}
{\renewcommand{\arraystretch}{1.3}
\begin{tabular}{| c | c | c | }
\hline
	$J^P$ & Wavefunction & $E_{\mathrm{exp}}$(MeV) \\ \hline
	$0^+_1$ & $\psi_{T0}^+ \ket{ 0, 0 }$ & 0 \\
	$0^+_2$ & $\psi_{T2}^+ \ket{ 0, 0 }$ & 6.0 \\
	$2^+_1$ & $\frac{1}{\sqrt{8}}(u_1^+-v_1^+)\left(\ket{2,2} + \ket{2,-2}\right) - \frac{\sqrt{3}}{2}(u_1^++v_1^+)\ket{2,0}$ & 6.9\\
	$3^-_1$ & $\psi_{S0}^-\frac{1}{\sqrt{2}}\left(\ket{3,2} - \ket{3,-2}\right)$ & 6.1 \\
	$4^+_1$ & $ \sqrt{\frac{5}{24}}\psi_{T0}^+\left( \ket{4,4} + \sqrt{\frac{14}{5}}\ket{4,0} + \ket{4,-4} \right) $ & 10.4 \\
	$4^+_2$ & $\sqrt{\frac{7}{32}}(u_1^++v_1^+)(\ket{4,4} + \ket{4,-4}) - \frac{1}{\sqrt{8}}(u_1^+-v_1^+)(\ket{4,2} + \ket{4,-2})$ & 11.1 \\
	&$-\frac{\sqrt{5}}{4}(u_1^++v_1^+)\ket{4,0}$ & \\
	\hline

\end{tabular}}
\caption{The wavefunctions, in terms of vibrational wavefunctions and spin states, for each of the states considered in this paper. Each model state is identified with an experimental state, whose energy is also tabulated.  We suppress the $J_3$ label for ease of reading.} 
\end{center}
\end{table}

To help analyse and compare results, it is helpful to introduce the idealised rigid body as a benchmark model. Here, the nucleus is described as four alpha-particles that form a rigid geometric shape which is allowed to rotate as a whole. The rotational motion is quantised and leads to rotational bands. Different shapes can lead to different rotational bands. For Oxygen-16, the $0_1^+, 3_1^-, 4_1^+$ states are understood as the rotational band of a tetrahedron while the $0_2^+, 2_1^+, 4_2^+$ states arise as the rotational band of a square (or possibly a chain \cite{Bau}, though this idea was recently dismissed experimentally \cite{nochain}). The most important parameter in this model is the ratio of the separation between the particles which form the tetrahedron $r_t$ and the separation between the particles which form the square $r_s$. We take
\begin{equation}
\frac{r_s}{r_t} = 1.5 \, ,
\end{equation}
and then fix $r_t$ to match the $B(E3; 3_1^- \to 0_1^+)$ transition. This is probably not a realistic model, but displays some important features that highlight the physics at play.

\subsection{Results}

The electromagnetic transition rates for our model, the rigid body model, the ab initio calculation \cite{abin} and the ACM \cite{BI3} are displayed in Table \ref{oxy16results} . They should be compared to the experimental data, which is also tabulated.

\begin{table}
	\begin{centering}
		\begin{tabular}{|c|c|c|c|c|c|}
			\hline 
			$B\left(El,i\rightarrow f\right)$ & our model & rigid body & ``rescaled"  & ACM \cite{BI2} & experiment \\
			& &model & ab initio \cite{abin} & &$\left[e^{2}\text{fm}^{2l}\right]$\cite{exp16}  \\
			\hline 
			$B\left(E3,3_{1}^{-}\rightarrow0_{1}^{+}\right)$ & 205 & 205 & & 215 & $205 \pm 11$ \\
			$B\left(E4,4_{1}^{+}\rightarrow0_{1}^{+}\right)$ & 320 & 633 & &  425 & $378 \pm 133$ \\
			
			$B\left(E6,6_{1}^{+}\rightarrow0_{1}^{+}\right)$ & 11263 & 23764 &  & 9626 &  \\

			$B\left(E1,2_{1}^{+}\rightarrow3_{1}^{-}\right)$ & 0 & 0 &  &  &  $ < 1.6 \times 10^{-5} $  \\
			$B\left(E1,4_{1}^{+}\rightarrow3_{1}^{-}\right)$ & 0 & 0  & &   & $< 1.2 \times 10^{-5}$ \\
			$B\left(E1,4_{2}^{+}\rightarrow3_{1}^{-}\right)$ & 0 &0 &  &   & $(2.4\pm 1) \times 10^{-5}$ \\
			
			$B\left(E2,2_{1}^{+}\rightarrow0_{1}^{+}\right)$ & 16 & 0 & $6.2 \pm 1.6$ & 26 &  $7.4 \pm 0.2$\\
			$B\left(E2,2_{1}^{+}\rightarrow0_{2}^{+}\right)$ & 22 & 70 & $46 \pm 8$ & 6 &  $65 \pm 7$ \\
			$B\left(E2,2_{1}^{-}\rightarrow3_{1}^{-}\right)$ & -- & 0 & & 10 & $13.4 \pm 3.8$ \\
			$B\left(E2,4_{1}^{+}\rightarrow2_{1}^{+}\right)$ & 13 & 0 & & 0 &  $146 \pm 17$  \\
			$B\left(E2,4_{2}^{+}\rightarrow2_{1}^{+}\right)$ & 7 & 100 & & 36 &  $2.4 \pm 0.7$ \\

			$B\left(E4,4_{1}^{+}\rightarrow0_{2}^{+}\right)$ & 24 & 0 &  &   &  \\
			$B\left(E4,4_{2}^{+}\rightarrow0_{1}^{+}\right)$ & 592 & 0 & &   &  \\
			$B\left(E4,4_{2}^{+}\rightarrow0_{2}^{+}\right)$ & 1632 & 8801 & &   &  \\
			\hline 
		\end{tabular}
		\par\end{centering}
	\caption{EM transition rates $B\left(El,i\rightarrow f\right)$ for Oxygen-16. We tabulate the results for the model described in this Section, the ab initio calculation and the Algebraic Cluster Model, as well as the available experimental data. All values are in units of $e^2$fm$^{2l}$.}
	\label{oxy16results}

\end{table}

The transition rates along the lowest lying band are in good agreement with experimental data in our model. These states are constructed from $\psi_{T0}^+$, which is concentrated at the tetrahedron. Hence, this result supports the idea that these states are tetrahedral in nature. The value for the $E6$ transition is close to the value from the ACM. This is to be expected, as the states have similar descriptions in both models.

The rigid body model highlights some important physics, though is an extreme approximation as can be seen from the enormous $E6$ transition. Since the square is more spread out than the tetrahedron, the square-like states (such as $0_2^+$, $2_1^+$ and $4_2^+$) have large transition rates between them. Similarly, the states in our vibrational model which contain significant square contributions lead to larger transition rates. For instance $B(E2; 2_1^+ \to 0_2^+) > B(E2; 2_1^+ \to 0_1^+)$, since $0_2^+$ is physically a mixture of the two shapes while $0_1^+$ contains little square contribution. This ordering is seen experimentally but the magnitudes of the transition rates are wrong in our model. For instance, the $B(E2,2_1^+ \to 0_2^+)$ is too small. This may be due to the approximations made in constructing the wavefunctions. We neglect the effect that a changing shape has on the structure of the wavefunction. This is because we take a constant moment of inertia tensor over the space of configurations. Hence, the $2_1^+$ wavefunction doesn't account for the fact that the square is much flatter than the tetrahedron. If we did account for this, the wavefunction would be more concentrated at the square and the transition rate would be enhanced. Note that the Carbon-12 calculation does account for this effect. To do the same calculation for the Oxygen-16 case, it would be necessary to solve the full Schr\"odinger equation on the 2-dimensional, 6-punctured sphere or develop a quantum graph model. This partially explains the discrepancy between the vibrational and rigid body models. The problem is even more pronounced in the $B(E2; 4_2^+ \to 2_1^+)$ transition rate. Naively, one would expect this to be large: physically, both states are square-like. As we can see from the rigid body model, this should lead to a large transition rate. But the transition rate is significantly diminished in our model, due to our approximations.

Although the rigid body model can generate large transition rates (which are seen in nature), it also predicts many erroneous zero results. This is easily understood: states can only decay along rotational bands. This is not seen in the experimental data, and suggests the model is too constrained. Similarly, the ACM predicts many small or zero results which are not in agreement with data. The vibrational model allows for greater overlap between wavefunctions and hence there are no zero results for any transitions, except the $E1$ transitions.  Unfortunately, the true amount of mixing is underestimated in all models.

There is one major discrepancy between all models and data. The $B(E2; 4_1^+ \to 2_1^+$) transition has a value of $(146\pm17)e^2$fm$^4$, while the rigid body, ACM and vibrational models give predictions of $0,0$ and $13$ respectively: at best an order of 10 too small. Such a large transition rate is very rare, so to find any possible explanation is worthwhile. One idea is that the $4_1^+$ state has been historically mis-characterised as a tetrahedral state. Suppose instead that the low lying $0_2^+,2_1^+,4_1^+$ band is a rotational band, of either the square or chain configurations. Then there is the following relationship for transition rates between states on the band
\begin{equation}
\frac{B(E2; 4_1^+ \to 2_1^+)}{ B(E2; 2_1^+ \to 0_2^+)} = \frac{10}{7} \approx 1.43 \, .
\end{equation}
In reality, the experimental ratio is
\begin{equation}
\frac{146 \pm 17}{65 \pm 7} = 2.25 \pm 0.5\, .
\end{equation}
This large ratio highlights the difficulty in describing the $B(E2; 4_1^+ \to 2_1^+)$ transition. The rigid body model, which should exaggerate this type of transition, still underestimates it. If one were to re-characterise the $4_1^+$ state as a rotational excitation of the square, the $4_2^+$ would then be interpreted as a tetrahedral state. The energy difference between the $4_1^+$ and $4_2^+$ states is only $0.74$ MeV, so their relabeling is reasonable on energetic grounds. As can be seen in Table 3, the $4_2^+$ state can still have a large E4 transition in the vibrational model, so this new interpretation may not spoil the positive results along the ground state band. To investigate further, one should improve the vibrational model to allow for a changing moment of inertia tensor, as described above. This should give more accurate results and will avoid underestimation.  Secondly, it may be worthwhile to redo the transition rate experiments.  These were last undertaken in the 1970s and early 1980s. Modern techniques would allow us to fill out Table 3 more fully. We are suggesting the spin 4 states may be mis-characterised, so having more information about the decay from the spin 4 states would be particularly useful. 

\section{Summary and further work}

Electromagnetic transition rates offer a wealth of information about the intrinsic structure of atomic nuclei. EM transitions help us to differentiate between the vast number of nuclear models on offer: shell model approaches, collective models, and the ACM to name a few. In this paper we developed a general formalism for computing EM transition rates within the framework of rotational-vibrational nuclear models.  

Within this formalism we calculated EM transition rates for two recently proposed models of Carbon-12 and Oxygen-16, which were inspired by nuclear dynamics in the Skyrme model. We found reasonable agreement with existing experimental data and highlighted important differences between our model's predictions and those of other models.

For Carbon-12 both our model and other models reproduce the existing data well. To differentiate the models more data is needed. We hope that this study provides fresh motivation to measure more EM transition rates for Carbon-12. The results for Oxygen-16 are less promising, for all models. We suggested that some discrepancies between experimental data and our model could be traced to our approximations. These may be improved by including a varying moment of inertia in our Schr\"odinger equation, or by developing a quantum graph model for the nucleus. No model comes close to full agreement with experimental data so there is still work to be done, even for these abundant nuclei. Further experimental data will help us to uncover their detailed structure.

We have focused on E transitions but M transitions are also seen experimentally. While E transitions depend on the charge density of the nucleus, the M transitions depend on the current density. These have been studied for Helium-3 and Hydrogen-3 within the Skyrme model \cite{Carson} but are not well understood in general.

\section{Acknowledgments}

We thank Nick Manton for helpful comments. CJH is supported by The Leverhulme Trust as an Early Careers Fellow. JIR was supported by an EPSRC studentship. This work has been partially supported by STFC consolidated grant ST/P000681/1.

\end{document}